\documentclass[aps,prd,twocolumn,showpacs,nofootinbib,amsmath,amssymb,amsfonts,showkeys]{revtex4}

\usepackage{graphicx}
\usepackage{bm}
\newcommand{\ra}{\rightarrow}

\begin{document}
\title{Do the cosmological observational data prefer phantom dark energy?}

\author{Bohdan Novosyadlyj}
 \email{novos@astro.franko.lviv.ua}
\author{Olga Sergijenko}
 \email{olka@astro.franko.lviv.ua}
\affiliation{Astronomical Observatory of 
Ivan Franko National University of Lviv, Kyryla i Methodia str., 8, Lviv, 79005, Ukraine}
\author{Ruth Durrer}
\email{ruth.durrer@unige.ch}
\affiliation{Universit\'e de Gen\`eve, D\'epartement de Physique Th\'eorique and CAP, 24 quai Ernest-Ansermet, CH-1211 Gen\`eve 4, Switzerland}
\author{Volodymyr Pelykh}
 \email{pelykh@iapmm.lviv.ua}
\affiliation{Ya. S. Pidstryhach Institute for Applied Problems of Mechanics and Mathematics, \\ Naukova str., 3-b, Lviv, 79060, Ukraine}

\date{\today}

\begin{abstract}
The dynamics of expansion and large scale structure formation of the Universe are analyzed for models with dark energy in the form of a 
phantom scalar field which initially mimics a $\Lambda$-term and evolves slowly to the Big Rip singularity. The discussed model of dark 
energy has three parameters -- the density and the equation of state parameter at the current epoch, $\Omega_{de}$ and $w_0$, and the 
asymptotic value of the equation of state parameter at $a\rightarrow\infty$, $c_a^2$. Their best-fit values are determined jointly with 
all other cosmological parameters by the MCMC method using  observational data on CMB anisotropies and polarization, SNe Ia luminosity 
distances, BAO measurements and more. Similar computations are carried out for  $\Lambda$CDM and a quintessence scalar field model of 
dark energy. It is shown that the current data slightly prefer the phantom model, but the differences in the maximum likelihoods are not 
statistically significant. It is also shown that the phantom dark energy with 
monotonically increasing density in future will cause the decay of large scale linear matter density perturbations due to the gravitational 
domination of dark energy perturbations long before the Big Rip singularity. 
\end{abstract}
\maketitle

\section{Introduction}

Among the large number of dark energy models the phantom model has probably the most exotic behavior
because it violates the null energy condition $\rho+p\ge0$. It was proposed first by R.~Caldwell~\cite{Caldwell2002} and 
A.~Starobinsky~\cite{Starobinsky2000} independently in 1999 as a possible explanation of the accelerated 
expansion of the Universe discovered a year earlier by two teams~\cite{Perlmutter1998,Riess1998}, 
measuring the  luminosity distance--redshift relation from SNe Ia in distant galaxies. It has been shown that this model of dark energy does 
not contradict the cosmological tests based on present data. The model was supported later by reconstruction of the equation of state parameter 
using new SN Ia data~\cite{Alam2004} and a fully consistent analysis of CMB, large scale structure and SN Ia data~\cite{Corasaniti2004}. 
The  'doomsday scenario' described in~\cite{Caldwell2003}, caused by the Big Rip singularity 
which is predicted by these models, surpasses the fantasy of science fiction writers. Since then, many authors have analyzed 
various aspects of phantom dark energy and confirmed its validity as candidate for dark energy~\cite{Carroll2003}. Moreover, 
WMAP~\cite{WMAP} data combined with either SNe Ia or BAO (Baryonic Acoustic Oscillations)  both prefer the phantom model of dark energy. 
This is why such models are the subject of active research in recent years: about a thousand papers devoted to phantom dark energy can be 
found in the publication databases.\\
\indent In Ref.~\cite{Caldwell2002} it has been shown that phantom dark energy can be modeled by a minimally coupled scalar field with 
classical Lagrangian apart from the  kinetic term which has the opposite sign. 
Other scalar field Lagrangians with non-canonical kinetic term leading to phantom-like properties of dark energy have 
been considered by different authors later. It turns out that some phantom models emerge effectively from the gravity sector of brane-world 
models~\cite{M-brane}, from superstring theory~\cite{Neupane2006}, from Brans-Dicke scalar-tensor gravity~\cite{Elizalde2004,Gannouji2006} 
and from quantum effects leading to violations of the weak energy condition on cosmological scales~\cite{Onemli2002,Onemli2004}. 
Some of these models have phantom properties only at the current stage of the evolution of the Universe but did not have them at early 
time or they lose this feature in the future\footnote{We do not consider here models crossing the phantom divide like e.g.  'quintom' dark 
energy, see Ref.~\cite{Feng2005}.}.\\
\indent It was shown by Carrol et al. (2003) and Cline et al. (2004) \cite{Carroll2003} that minimally coupled scalar fields with a linear negative 
kinetic term may cause a UV quantum instability of the vacuum manifesting itself in the production of pairs of ghosts, 
photons or gravitons as a consequence of the violation of the null energy condition\footnote{This can be prevented by introducing the squared kinetic term 
in the Lagrangian as in the ghost condensate model~\cite{Arkani2004} or by second derivatives of the scalar field as in the kinetic braiding scalar-tensor 
model~\cite{Deffayet2010}.}. For late type phantom scalar fields the produced ghosts typically carry low energy, so, their decay rates are strongly 
time-dilated. On the other hand, the time scale of this instability for phantom dark energy can be much larger than the cosmological one, making this 
effect unsuitable for constraining  the parameters of the model 
at the present level of observations. This is why in this paper we concentrate our attention on the classical properties of phantom scalar field 
models of dark energy and on possibilities to determine their parameters by comparison of predictions with available observational data.\\ 
\indent The main feature of phantom dark energy is its strongly negative equation of state parameter, $w_{de}<-1$, which implies an energy density 
increasing with time. If $w_{de}=const$, the energy density of such a field is zero at the Big Bang, it starts from "nothing", that is why 
it is dubbed ghost or phantom. In this paper we will show that this true phantom can arise as a special case of a dynamical scalar field 
with barotropic equation of state which we call PSF (for phantom scalar field with barotropic equation of state) in the following. 
In the general case it starts as vacuum energy or a cosmological constant with $w_{de}=-1$ and evolves to a lower value $w_{de}<-1$ at the 
current epoch and to a Big Rip singularity in the distant future. The PSF model of dark energy has three parameters, we determine their 
best-fit values jointly with other cosmological parameters using available observational data on CMB anisotropies and polarization, 
SNe Ia luminosity distances, BAO measurements and others. We compare 
the maximum likelihood of the studied PSF model with corresponding values for a  barotropic scalar field with positive kinetic term, which 
we call QSF (for quintessence scalar field with barotropic equation of state), and with $\Lambda$CDM for the same data sets.\\
\indent The paper is organized as follows. In Section~\ref{bckgr} we discuss the cosmological dynamics of phantom scalar fields with barotropic EoS. 
In Section~\ref{psf_instab} we analyze the gravitational instability of  PSF and its effects on structure formation. In Section~\ref{best-fit} 
we present the results of  an estimation of  PSF parameters and compare the goodness of fit of three types of models (PSF , QSF  and 
$\Lambda$CDM) for the same data sets. The conclusions are found in Section~\ref{conclusions}.  

\section{Evolution of phantom scalar field and expansion of the Universe}\label{bckgr}

\begin{figure*}
  \includegraphics[width=.45\textwidth]{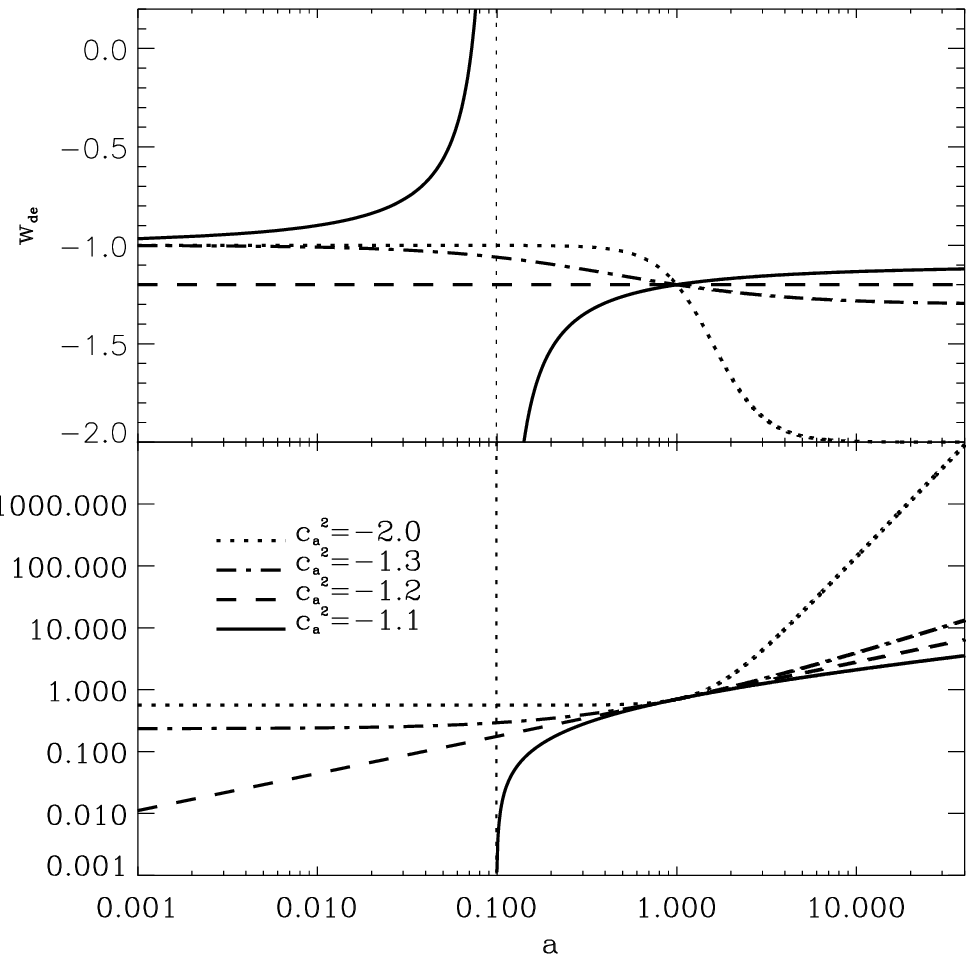}
  \includegraphics[width=.45\textwidth]{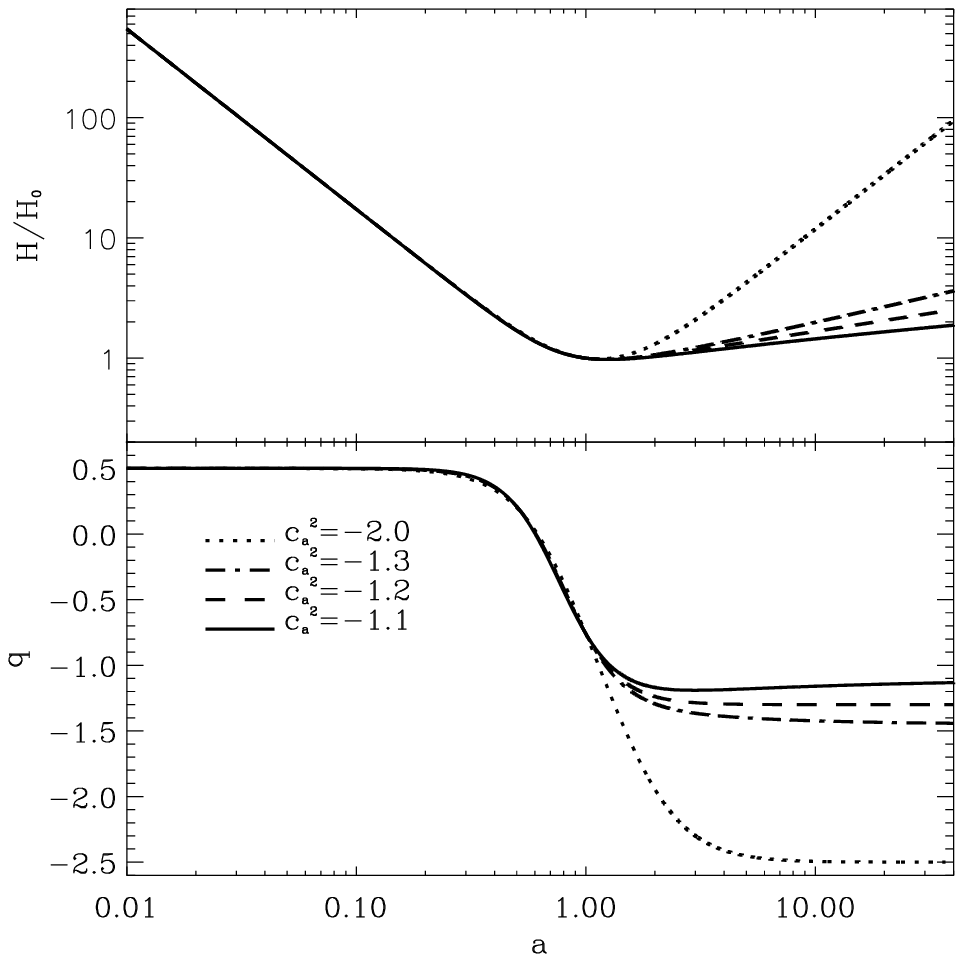}
  \caption{Left column: top panel -- the dependence of EoS parameter
    on the scale factor for barotropic phantom scalar field with $w_0$=-1.2 and
    different $c_a^2$ (-2.0, -1.3, -1.2, -1.1); bottom panel -- the
    dependence of the dark energy density (in the units of critical one
    at the current epoch) on the scale factor with the same EoS parameters.  Right column: top panel -- the
    dynamics of the expansion of the Universe with barotropic phantom scalar field:
    $H^2(a)$ (top panel) and $q(a)$ (bottom panel) for the same models as in the left panels.}
  \label{wrode_ph}
\end{figure*}

Let us analyze phantom dark energy from a single minimally coupled scalar field. Real scalar fields with classical Lagrangian $L=X-U(\phi)$, 
where $X=\phi_{,i}\phi^{,i}/2$ is the kinetic term, or with a tachyonic Lagrangian, $L=-U(\phi)\sqrt{1-2X}$, cannot be models of phantom dark energy 
since the kinetic term is positive. The simplest Lagrangian allowing $w<-1$ is that of a  classical scalar field with a kinetic term of opposite sign:
 \begin{equation}
  L_{de}=-X-U(\phi).\label{L_ph1}
\end{equation}
This has been proposed in \cite{Caldwell2002} and~\cite{Starobinsky2000} and it is inherent from spacelike brane constructions in string theory~\cite{S-brane}.
In this case the energy density and pressure are the following linear combinations of $X$ and $U$: 
\begin{eqnarray}
  \rho_{de}=-X+{U}(\phi), \quad  p_{de}=-X-{U}(\phi). \label{rho_XU_ph1} 
\end{eqnarray}
The EoS parameter
\begin{eqnarray}
  w_{de}=\frac{-X-U}{-X+U}\label{w_XU_ph1} 
\end{eqnarray}
for positive values of $X$ and $U$ is less than $-1$. In order to explain the accelerated expansion of the Universe at the current epoch ($q_0<0$) 
the phantom scalar field must satisfy two conditions:
\begin{eqnarray}
 {\rm a)}\,\, 0<X^{(0)}<U^{(0)}, \quad  {\rm b)}\,\, U^{(0)}+2X^{(0)}>\rho_m^{(0)}/2.
\end{eqnarray}
 
We specify the scalar field model of dark energy by the condition $\dot{p}_{de}/\dot{\rho}_{de}=c_a^2=const$. This  is equivalent of the generalized 
linear barotropic equation of state $p_{de}=c_a^2\rho_{de}+C$, where $C$ is a constant. Representing the equation of state as $p_{de}=w_{de}\rho_{de}$ 
from the energy-momentum conservation law in the Friedmann-Robertson-Walker metric with the scale factor $a(t)$ the analytic solutions for $w_{de}(a)$ 
and $\rho_{de}(a)$ have been obtained in \cite{Novosyadlyj2010}:
\begin{eqnarray}
 w_{de}(a)=\frac{(1+c_a^2)(1+w_0)}{1+w_0-(w_0-c_a^2)a^{3(1+c_a^2)}}-1,\label{w}\\
\rho_{de}=\rho_{de}^{(0)}\frac{(1+w_0)a^{-3(1+c_a^2)}+c_a^2-w_0}{1+c_a^2},\label{rho}
\end{eqnarray}
where $w_0\equiv w_{de}(a=1)$ is the initial condition for integration of the differential equation 
$w_{de}'=3a^{-1}(1+w_{de})(w_{de}-c_a^2)$ (a prime denotes the derivative with respect to the scale factor $a$).
This simplifies significantly the analysis of cosmological consequences of this field. 
From (\ref{w}) it follows that at the current epoch ($a=1$) $w_{de}=w_0$. For a phantom scalar field (taking into account 
that $a\rho_{de}'=-3\rho_{de}(1+w_{de})>0$) the value of $w_0$ must be less than -1. The additional condition $w_{de}\le-1$ 
for any $a$ gives immediately the constraint $c_a^2\le w_0$ and $w_{de}(\infty)=c_a^2$. If $c_a^2>w_0$ the phantom divide 
($w_{de}=-1$) is crossed in the past, the energy density of scalar field changes sign and the equation of state parameter 
diverges at $a_s$ given by $a_s^{3(1+c_a^2)}=-(1+w_0)/(c_a^2-w_0)$. Here we exclude such models from our considerations. 
In Fig.~\ref{wrode_ph} (left column) the evolution of $w_{de}$ and $\rho_{de}$ is shown for some values of $w_0$ and $c_a^2$. 
One can see that the energy density of phantom barotropic scalar field increases for all combinations of $w_0<-1$ and $c_a^2<-1$. 
\begin{figure*}
  \includegraphics[width=.43\textwidth]{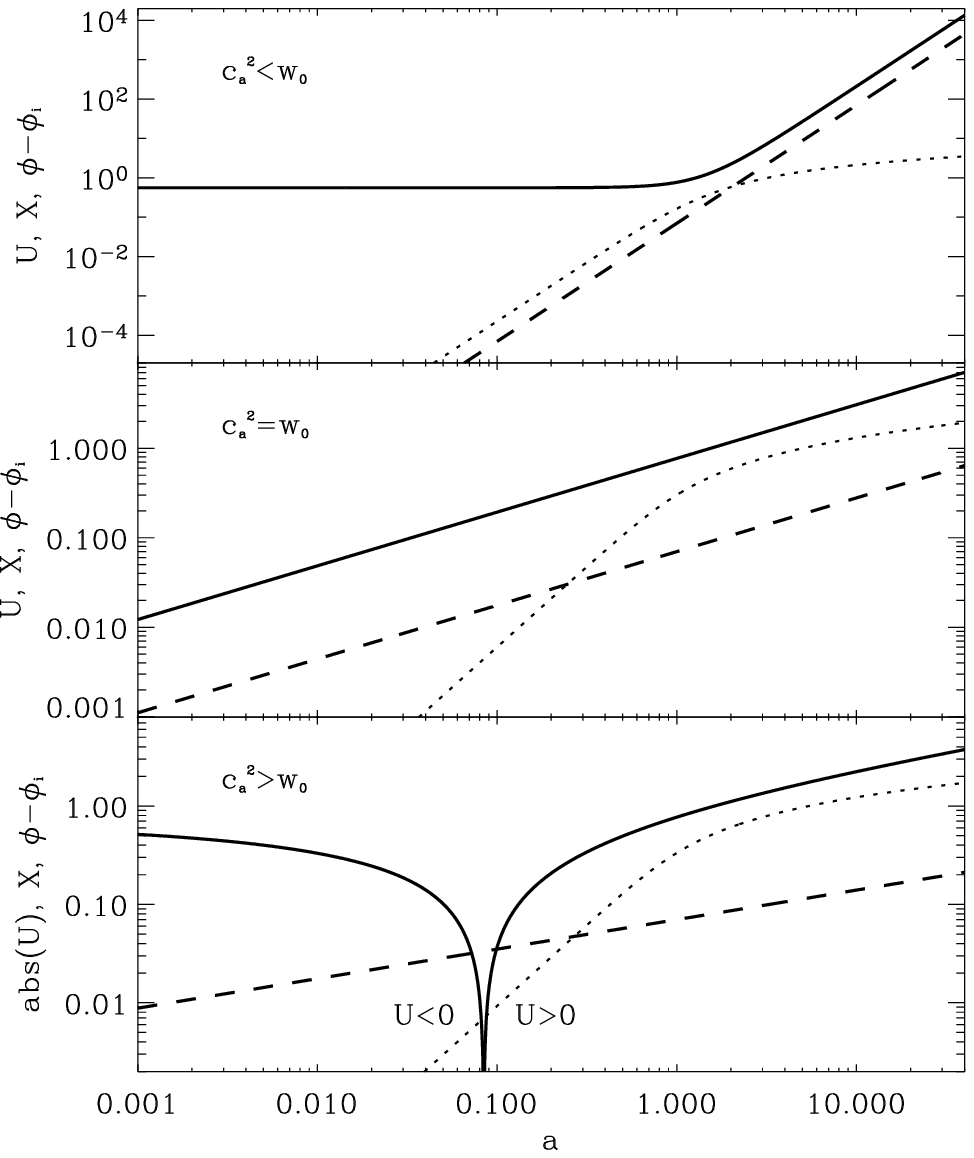}
  \includegraphics[width=.43\textwidth]{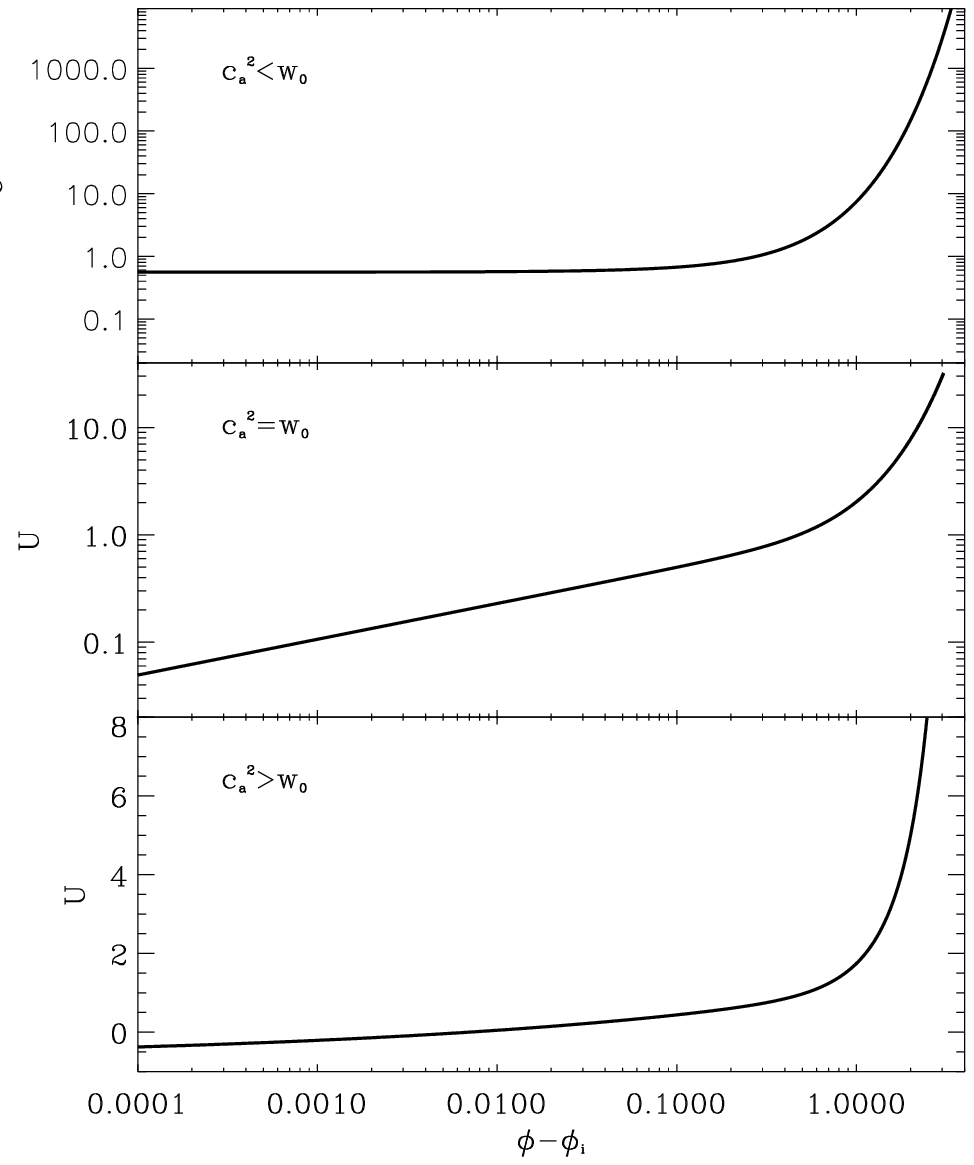}
  \caption{Right panels: The dependences of potentials $U$ (solid line), field variable $\phi-\phi_i$ (dotted) and the kinetic term $X$ (dashed) 
as functions of the scale factor $a$ for PSF with Lagrangian (\ref{L_ph1}) and barotropic EoS (\ref{w}) with different relations between $w_0$ 
and $c_a^2$ ($c_a^2=-2.0,\,-1.2,\,-1.1$ from top to bottom, $w_0=-1.2$ for all). Left panel: The dependences of potentials $U$ on $(\phi-\phi_i)$ 
for the same PSF models. In all except the  bottom right panels the dependences are in $\log{}-\log{}$ scales, in the right bottom panel  we 
use$\log{}$-lin scales. The potential and the kinetic term are in the units of current critical energy density, $3c^2H_0^2/8\pi G$, the field variable 
is in units of $\sqrt{3c^2/8\pi G}$. In the right panels field evolves from left to right.}
\label{U_X_ph1}
\end{figure*}

The quintessence sector of barotropic scalar fields  ($-1<w_0\le-1/3$, $-1<c_a^2\le0$), denoted QSF, 
where the density of the scalar field decreases with 
expansion more slowly than the density of matter, has been studied in detail in our previous
papers~\cite{Novosyadlyj2010,Novosyadlyj2011,Sergijenko2011}. This case requires $c_a^2>w_0$ in order for the dark energy to remain positive. 
A comparison of the models is given in Table~\ref{t:QP}.
\begin{table}[ht]
\caption{\label{t:QP} The asymptotic behavior of quintessence (QSF) and phantom (PSF) barotropic scalar field models.}
\begin{tabular}{c|c|c|c|c} \hline
 & $a\ra 0$ & $a=1$ & $a\ra \infty$ & Future of the Universe\\
 \hline
 {\bf QSF} & & & & $-1\le w_0<-1/3$ \\
  &&&& $w_0<c_a^2\le 0\,:$  \\
 $w_{de}$   & $c_a^2$ & $w_0$ & $-1$ & eternal inflation \\
 $\rho_{de}$ & $\infty$ & $\rho_{de}^{(0)}$ & $\rho_{de}^{(0)}\frac{c_a^2-w_0}{1+c_a^2}$ & $-1<c_a^2<w_0<-1/3\,:$  \\
&&&& Big Crunch singularity\\
  \hline
 {\bf PSF} & & & & $c_a^2<w_0< -1$\\
 &&&& Big Rip singularity\\
 $w_{de}$   & $-1$ & $w_0$ &  $c_a^2$ & \\
 $\rho_{de}$ & $\rho_{de}^{(0)}\frac{c_a^2-w_0}{1+c_a^2}$ & $\rho_{de}^{(0)}$&  $\infty$ & \\
  \hline
\end{tabular}
\end{table}

We consider a multicomponent model of the Universe filled with non-relativistic particles (cold dark matter and baryons), relativistic particles 
(thermal electromagnetic radiation and massless neutrino) and a phantom scalar field as described above, termed PSF+CDM. The background Universe 
is assumed to be spatially flat, homogeneous and isotropic with Friedmann-Robertson-Walker (FRW) metric,
$$ds^2=g_{ij} dx^i dx^j =a^2(\eta)(d\eta^2-\delta_{\alpha\beta} dx^{\alpha}dx^{\beta}),$$ 
where $\eta$ is the conformal time defined by $dt=a(\eta)d\eta$ and $a(\eta)$ is the scale factor, normalized to 1 at the current epoch 
(here and below we put $c=1$). Latin indices \textit{i, j,...} run from 0 to 3 and the Greek ones are used for the spatial part of the metric:
 $\alpha,\,\beta,..=1,2,3$. The dynamics of the expansion of the Universe can be deduced from the Einstein equations
\begin{eqnarray}
R_{ij}-{\frac{1}{2}}g_{ij}R=8\pi G \left(T_{ij}^{(m)}+T_{ij}^{(r)}+T_{ij}^{(de)}\right),
\label{E_eq}
\end{eqnarray}
where $R_{ij}$ is the Ricci tensor and $T_{ij}^{(m)},\,T_{ij}^{(r)},\,T_{ij}^{(de)}$ are the energy-momentum tensors of non-relativistic matter (m), 
relativistic matter (r), and dark energy (de) respectively. Assuming that the interaction between these components is only gravitational, each of 
them satisfies the differential energy-momentum conservation law separately: $T^{i\,(N)}_{j;i}=0$. Einstein's equations together with energy 
conservation lead to the Friedman equations, which describe the rate and acceleration of the expansion of the Universe:
\begin{eqnarray}
 H&=&H_0\sqrt{\Omega_r/a^{4}+\Omega_m/a^{3}+\Omega_{de}f(a)}, \label{H}\\
 q&=&\frac{1}{2}\frac{2\Omega_r/a^{4}+\Omega_m/a^{3}+(1+3w_{de})\Omega_{de}f(a)}
{\Omega_r/a^{4}+\Omega_m/a^{3}+\Omega_{de}f(a)},\label{q}
\end{eqnarray}
where $f(a)=\rho_{de}(a)/\rho_{de}(1)$. Here $H\equiv\dot{a}/{a^2}$ is the Hubble parameter (expansion rate), 
$q\equiv-\left(a\ddot{a}/\dot{a}^2-1\right)$ is the deceleration parameter and an overdot denotes derivative w.r.t.
conformal time $\eta$. Eqs.~(\ref{H})-(\ref{q}) completely describe the dynamics of expansion of the homogeneous and isotropic Universe. 
In the past it was dominated by radiation and matter, in the distant future it is dominated  by the phantom scalar field, as  shown in the 
Fig. \ref{wrode_ph}. One can see that the rate of expansion $H$ (top right panel) decreases when the deceleration parameter is $q>0$ and starts 
to increase when the phantom component begins to dominate. In the case of a quintessence scalar field with barotropic EoS $H$ decreases always, 
asymptotically approaching a constant value in the case of $c_a^2>w_0$ leading to late inflation \cite{Novosyadlyj2010}.
\begin{figure}[ht]
\includegraphics[width=.47\textwidth]{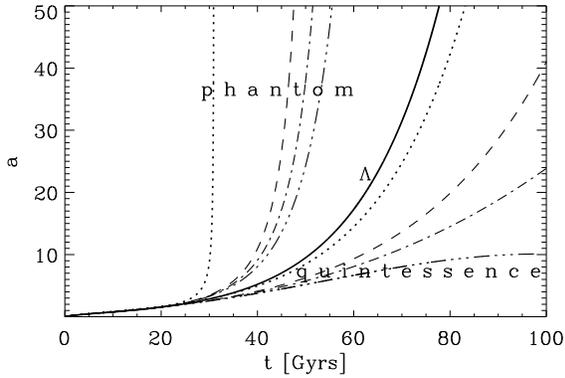}
\caption{Dependences of the scale factor on time, $a(t)$, for cosmological models with quintessence/phantom scalar fields with $w_0=-1\pm0.2$ 
and $c_a^2=-1\pm1$ (dotted line), $-1\pm0.3$ (dashed one), $-1\pm0.2$ (dash-dotted), $-1\pm0.1$ (dash-three-dotted). 
The upper sign is for QSF, the lower one for PSF. For the $\Lambda$CDM model $a(t)$ is shown by thick solid line. Note also that the limiting 
behavior for PSF with $1+c_a^2 \nearrow 0$  does tend to the cosmological constant behavior while $1+c_a^2 \nearrow 0$ does not. In all models 
$\Omega_m=0.3$, $\Omega_{de}=0.7$, $H_0=70$ km/s$\cdot$Mpc.}
\label{ah_ph}
\end{figure} 

Assuming a barotropic equation of state the field variable, the potential and the kinetic term can be obtained in terms of the
scale factor and the three model parameters $(\rho_{de}^{(0)}, w_0, c_a^2) $ as follows:
\begin{eqnarray}
  &&\phi(a)-\phi_{i}=\pm\sqrt{-(1+w_0)\rho_{de}^{(0)}}\int_0^a\frac{da'}{a'^{(\frac{5}{2}+\frac{3}{2}c_a^2)}H(a')},\nonumber\\
  &&U(a)=\frac{(1-c_a^2)(1+w_0)a^{-3(1+c_a^2)}+2(c_a^2-w_0)}{2(1+c_a^2)}\rho_{de}^{(0)}, \nonumber\\
  &&X(a)=-\frac{1+w_0}{2}a^{-3(1+c_a^2)}\rho_{de}^{(0)}.\label{U_ph1}
\end{eqnarray}
The phantom barotropic scalar field ($w_0<-1$, $c_a^2<-1$) has real field values if the current dark energy density is non-negative. Its kinetic term 
$X(a)$ is positive for all $a$, the potential $U(a)$ is positive as long as $c_a^2\le w_0$. If $w_0<c_a^2<-1$, $U(a)$ starts from the negative value 
$(c_a^2-w_0)\rho_{de}^{(0)}/(1+c_a^2)$ at $a=0$, changes the sign at $a_{(\rho=0)}=[2(w_0-c_a^2)/(1-c_a^2)(1+w_0)]^{-\frac{1}{3(1+c_a^2)}}$, which 
for the phantom case is always $\le 1$. In any case $U(a)$ increases with $a$. This distinguishes the phantom scalar field from the quintessence 
field. The evolution of the potential, $U(a)$, field variable\footnote{For definiteness and graphic representation we put the undetectable initial 
value $\phi_i$ equal to zero.}, $\phi(a)$, and kinetic term, $X(a)$, for models with $c_a^2<w_0<-1$, $c_a^2=w_0<-1$ and $w_0<c_a^2<-1$ are shown 
in Fig.~\ref{U_X_ph1}. In the right panels the reconstructed potential $U(\phi-\phi_i)$ is also shown. The accelerated expansion of the 
Universe is caused by the rolling  of the field up to the maximum of its potential, inversely to the case of quintessence scalar field. 
The energy density and pressure are smooth monotonic  functions of $a$ for all relations 
between $c_a^2<-1$ and $w_0<-1$, while $w_{de}$ has a discontinuity of the second kind in the case $w_0<c_a^2<-1$, when the scalar field energy 
density passes through zero (see Fig. \ref{wrode_ph}, left bottom panel). 

Another difference between  PSF and QSF is in their asymptotic behavior: PSF mimics a cosmological constant at the Big Bang for any $c_a^2<-1$ ($w_{de}$ 
tends to -1 when $a$ tends to 0), while QSF does this at $a\rightarrow\infty$ in the case of decreasing EoS parameter, see Table~\ref{t:QP}. 
PSF always starts as cosmological constant with $\rho_{de}(a=0)=\rho_{de}^{(0)}(c_a^2-w_0)/(1+c_a^2)$, which is positive for $c_a^2<w_0$ and negative 
if $w_0<c_a^2<-1$. This property distinguishes the barotropic phantom scalar field  from the ``standard'' phantom dark energy discussed 
in~\cite{Caldwell2002,Caldwell2003}, where the density starts from zero at $a=0$. For PSF only the special case with $c_a^2=w_0$ has this behavior. 

In the future, when $a\gg1$, the energy density of PSF increases as $\rho_{de}(a)\propto (1+w_0)/(1+c_a^2)\rho^{(0)}_{de}a^{-3(1+c_a^2)}$ 
while $w_{de}$ tends to $c_a^2$ (eqs. (\ref{w}) and (\ref{rho})). The repulsion properties of PSF increase and in finite time they reach and 
outmatch first the gravitational force, then electromagnetic forces and finally strong interactions. All bound structures in the 
Universe -- galaxies, stars, planets, atoms and protons -- will be ripped apart in finite time. This singularity is dubbed the 
Big Rip~\cite{Caldwell2003} and the moment when it happens can be estimated from the time dependence of the scale factor:
\begin{equation}
  t(a) =\int_0^a\frac{da'}{a'H(a')},
  \label{a(t)}
\end{equation}
which can be computed numerically using (\ref{H}) for any cosmological model and parameters of scalar field with barotropic EoS. In Fig. \ref{ah_ph} 
we present the time dependences of scale factors, $a(t)$, for cosmological models with PSF with the same parameters as in Fig.~\ref{wrode_ph}. For 
comparison we show also 
$a(t)$ for the $\Lambda$CDM and QSF+CDM models with symmetrical values of $w_0$ and $c_a^2$ relative to the phantom divide line. Phantom regime for 
$a(t)$  lies above the $a(t)$-curve for $\Lambda$CDM with the same cosmological parameters, while quintessence range is below. At $a\gg1$, when 
radiation and matter terms in (\ref{H}) can be neglected, we obtain the following approximate analytic expression for $a(t)$:
\begin{equation}
 a(t)\approx\left[\frac{3}{2}H_0(1+c_a^2)\sqrt{\frac{(1+w_0)\Omega_{de}}{1+c_a^2}}(t-t_0)+1\right]^{\frac{2}{3(1+c_a^2)}}\label{a_ph}.
\end{equation}
This shows that $a\rightarrow\infty$ is reached within finite time
\begin{equation}
 t_{BR}-t_0\approx \frac{2}{3}\frac{1}{H_0}\frac{1}{|1+c_a^2|}\sqrt{\frac{1+c_a^2}{(1+w_0)\Omega_{de}}}\label{t_br},
\end{equation}
which is noted as the time of Big Rip \cite{Starobinsky2000,Caldwell2003}. This super-fast expansion leads also to freezing of the particle horizon 
$r_p$ at some $r_p^{max}$ and 
to a decay of the event horizon $r_e$ to zero at $t\rightarrow t_{BR}$. (We denote expansion which is faster than exponential by 
'super-fast' expansion.) More precisely, in comoving coordinates these quantities behave  as follows:
\begin{eqnarray*}
r_{p}(t)= \int^{a}_{0}\frac{da'}{a'^{2}H(a')},\quad r_{e}(t)= \int^{\infty}_{a}\frac{da'}{a'^{2}H(a')} .
\end{eqnarray*}
Starting from $a_f\gg1$, when the matter component in (\ref{H}) can be neglected, the integral for $r_p$ from $a_f$ to $a\gg a_f$ can be computed 
analytically,
\begin{eqnarray*}
r_{p}(t) &=& r_p(t_f) + I(a_f,a), \\  I(a_f,a) &=& \hspace{-1.2mm}
\frac{2}{(1+3c_a^2)H_0}\sqrt{\frac{1+c_a^2}{(1+w_0)\Omega_{de}}}
\left(\!a_f^{\frac{(1+3c_a^2)}{2}}\hspace{-2.3mm} - a^{\frac{(1+3c_a^2)}{2}}\!\right),
\end{eqnarray*}
$ I(a_f,a)$ tends to 0 when $a_f\rightarrow\infty$. The finite time within which the singularity is reached, the freezing of the particle horizon 
and the decay of the  event horizon suggest that the dynamics of expansion of the Universe dominated by phantom dark energy is like the free fall 
into a Schwarzschild black hole.

The positive energy density of PSF becomes infinite in finite time (\ref{t_br}), overcoming all other forms of matter. The phantom scalar field dark 
energy rips apart first clusters of galaxies, later the Milky Way and other galaxies, then the solar system, a bit later the Sun and stars, the Earth 
and finally ``the molecules, atoms, nuclei and nucleons, which we are composed of, before the death of the Universe in a Big Rip'' (see Table 1 in 
\cite{Caldwell2003}). Will this be the end of Everything? Maybe this will be the beginning of new worlds -- if PSF reaches the Planck density, quantum 
fluctuations or interaction of the field with particles (the phenomenon of confinement) will lead to the inflation in some regions at Planck scales. In  
Ref.~\cite{Elizalde2004} it has been demonstrated that in a phantom Big Rip  quantum gravity effects might drastically change the future of our Universe, 
removing the singularity in a quite natural way.

\section{Gravitation instability of the PSF and large scale structure formation}\label{psf_instab}

Before determining the PSF parameters let us discuss briefly the gravitational instability of such a scalar field and its impact 
on matter clustering. The complete system of evolution equations for cosmological perturbations of cold dark matter, baryons, 
massless and massive neutrinos as well as radiation based on Einstein's equation, conservation laws and Boltzmann equations is 
presented in \cite{Ma1995,Durrer2008}.  Accurate line of sight integration can be performed using publicly available codes 
like CAMB \cite{camb} and CLASS \cite{class}. Here we use CAMB modified to include the  expressions (\ref{w}), (\ref{rho}), (\ref{H}) 
presented above and evolution  equations for scalar field perturbations \cite{Novosyadlyj2010}. 
In the general case the scalar field pressure perturbation includes in addition to the adiabatic component $\delta p^{(ad)}$ 
also a non-adiabatic mode $\delta p^{(nad)}$, $\delta p_{(de)}=c_a^2\delta \rho_{(de)}+\delta p_{(de)}^{(nad)}$, which is interpreted 
as the manifestation of intrinsic entropy of the scalar field (see \cite{Novosyadlyj2010} and references therein).
In the dark energy rest-frame the total pressure perturbation is expressed as $\delta p_{de}=c_s^2\delta\rho_{de}$, 
where the effective sound speed\footnote{The terms ``pressure``, ``entropy``, ``effective sound speed'' and 
``adiabatic sound speed'' of dark energy are used in the literature to denote dark energy properties which correspond only 
formally to the corresponding thermodynamical variables.} for the scalar field with given Lagrangian can be calculated as
$c_s^2\equiv p,_{X}/\rho,_{X}=L,_{X}/(2XL,_{XX}+L,_{X})$, first proposed in \cite{Garriga1999}. For the Lagrangian (\ref{L_ph1}) $c_s^2=1$ 
and the effective speed of sound (propagation of the perturbations) is equal to the speed of light. 
To understand the gravitational instability of  PSF and its impact on  large scale structure formation in the matter and dark energy 
dominated epochs it is sufficient to analyze the following set of differential equations:
\begin{eqnarray}
&&\dot{\delta}_{de}+3(c_s^2-w_{de})aH\delta_{de}+(1+w_{de})\frac{\dot{h}}{2}\nonumber\\ 
&&+(1+w_{de})\left[k+9a^2H^2\frac{c_s^2-c_a^2}{k}\right]V_{de}=0, \label{d_de}\\
&&\dot{V}_{de}+aH(1-3c_s^2)V_{de}-\frac{c_s^2k}{1+w_{de}}\delta_{de}=0,\label{V_de}
\end{eqnarray}
\begin{eqnarray}
&&\dot{\delta}_m=-\frac{1}{2}\dot{h}, \label{d_dm}\\
&&\hspace{-0.3cm}\ddot{h}+\frac{\dot{a}}{a}\dot{h}=-8\pi G a^2(\rho_m\delta_m+(1+3w_{de})\rho_{de}\delta_{de}),\label{h}
\end{eqnarray}
describing the evolution of density perturbations of dark energy $\delta_{de}\equiv \delta\rho_{de}/\rho_{de}$ and matter 
$\delta_m\equiv (\delta\rho_b+\delta\rho_{cdm})/(\rho_b+\rho_{cdm})$, velocity perturbation $V_{de}$ of dark energy and 
the evolution of metric perturbations $h\equiv h^i_i$ in the synchronous co-moving to dark matter gauge. Here as above an 
overdot denotes the derivative w.r.t. conformal time $\eta$. Adiabatic initial conditions for matter components can be 
found in \cite{Ma1995} and the initial conditions for the early time, when the scalar field is subdominant~\cite{Novosyadlyj2010} are as follows 
\begin{eqnarray}
&&\hspace{-5mm}\delta_{de}^{(in)}=-\frac{(4-3c_s^2)(1+w_{de})}{8+6c_s^2-12w_{de}+9c_s^2(w_{de}-c_a^2)}h^{(in)},\nonumber \\
&&\hspace{-6mm}V_{de}^{(in)}=-\frac{c_s^2k\eta_{in}}{8+6c_s^2-12w_{de}+9c_s^2(w_{de}-c_a^2)}h^{(in)}, \label{init} \\
&&\hspace{-5.2mm}\delta_m^{(in)}=-\frac{1}{2}h^{(in)}. \nonumber
\end{eqnarray}

\begin{figure}[ht] 
  \includegraphics[width=0.45\textwidth]{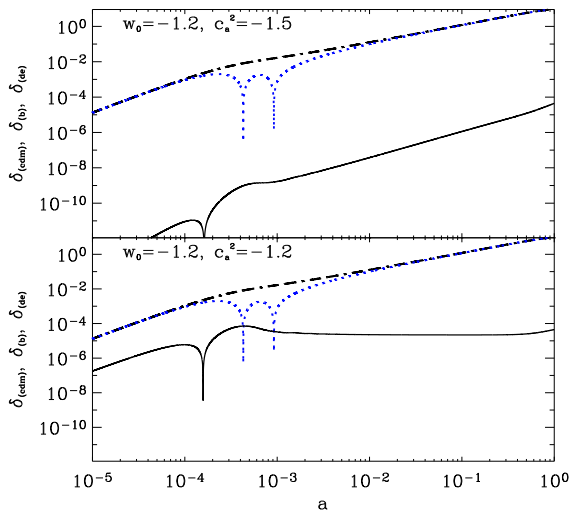}
  \caption{The evolution of the Fourier amplitude ($k=0.05$ Mpc$^{-1}$) of density perturbations for cold dark matter (dashed line), 
baryonic matter (dotted) and PSF (solid) for $c_a^2<w_0$ (top panel) and for $c_a^2=w_0$ (bottom panel).}
  \label{ddeb_p12_05}
\end{figure}

For positive matter density perturbation\footnote{In the early Universe for superhorizon scales we include also the relativistic components, 
which dominate the energy density.} ($\delta_m>0$) at $\eta_{in}$, the gravitational potential is negative $h<0$ and the dark energy density 
perturbation has the opposite sign ($\delta_{de}<0$) for any $w_0,\,c_a^2<-1$ and $c_s^2>0$. The absolute values of their amplitudes increase 
$\propto a$ on super-horizon scales, but  the density perturbations of the phantom scalar field change sign and decay after entering the 
horizon at $\eta\approx k^{-1}$. This is shown in Fig. \ref{ddeb_p12_05}, where the evolution of Fourier mode $k=0.05$ Mpc$^{-1}$ of density 
perturbations for dark matter, baryons and phantom scalar field is presented for 2 cases: $c_a^2<w_0$ and $c_a^2=w_0$.  

Note also, that PSF perturbations in general do not obey a barotropic equation of state even if the background does. 
But as we see here, they never become very relevant, we shall therefore not stress this any further.
 
In the case of $c_a^2=w_0$ the absolute value of the initial amplitude of $\delta_{de}$ is larger than for $c_a^2<w_0$, but in both cases 
during the structure formation and at present the amplitude of $\delta_{de}$ is significantly lower than $\delta_m$. This implies that the 
perturbations of minimally coupled scalar fields with initial conditions (\ref{init}) do not significantly affect structure formation. 

\begin{figure}[ht]
  \includegraphics[width=0.45\textwidth]{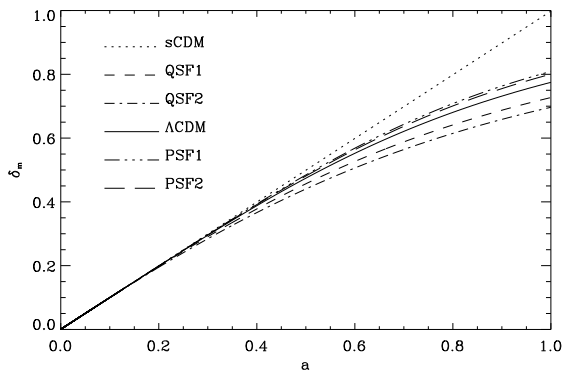}
  \caption{The evolution of matter density perturbations from the Dark Ages to the present in sCDM, 
  $\Lambda$CDM, 
QSF+CDM (1: $w_0=-0.8$, $c_a^2=-0.8$; 2: $w_0=-0.8$, $c_a^2=-0.5$) and PSF+CDM (1: $w_0=-1.2$, $c_a^2=-1.2$; 2: $w_0=-1.2$, $c_a^2=-1.5$) models. 
Amplitudes are normalized to $0.1$ at $z=10$ ($a=0.1$). In models with dark energy $\Omega_m=0.3$, $\Omega_{de}=0.7$.}
  \label{fig_dm}
\end{figure}
\begin{figure*}[ht]
\includegraphics[width=0.45\textwidth]{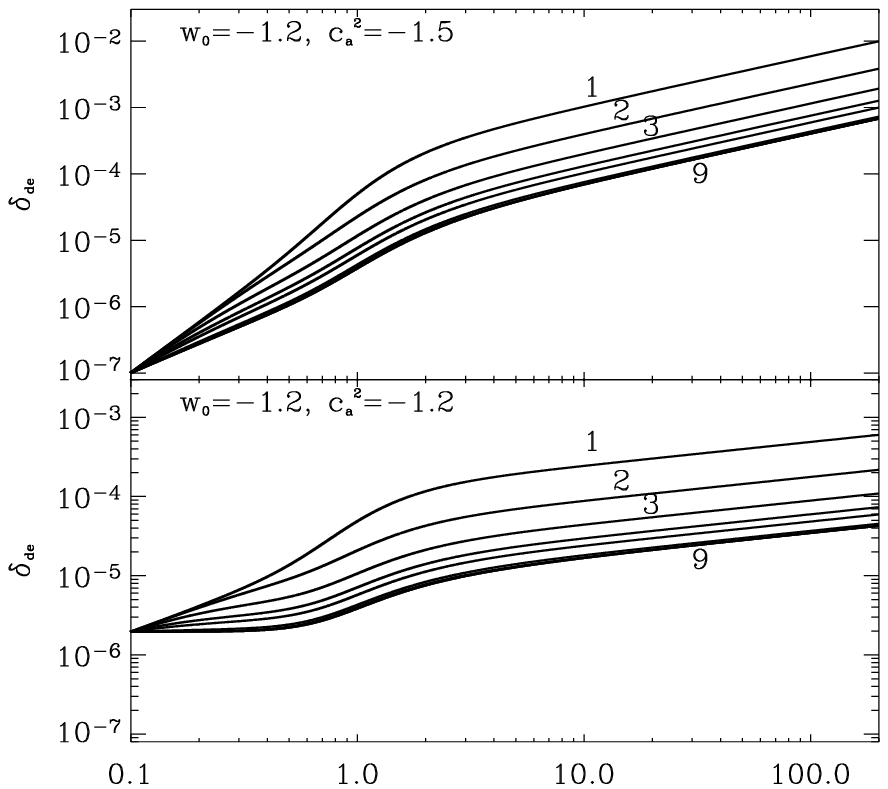}
\includegraphics[width=0.45\textwidth]{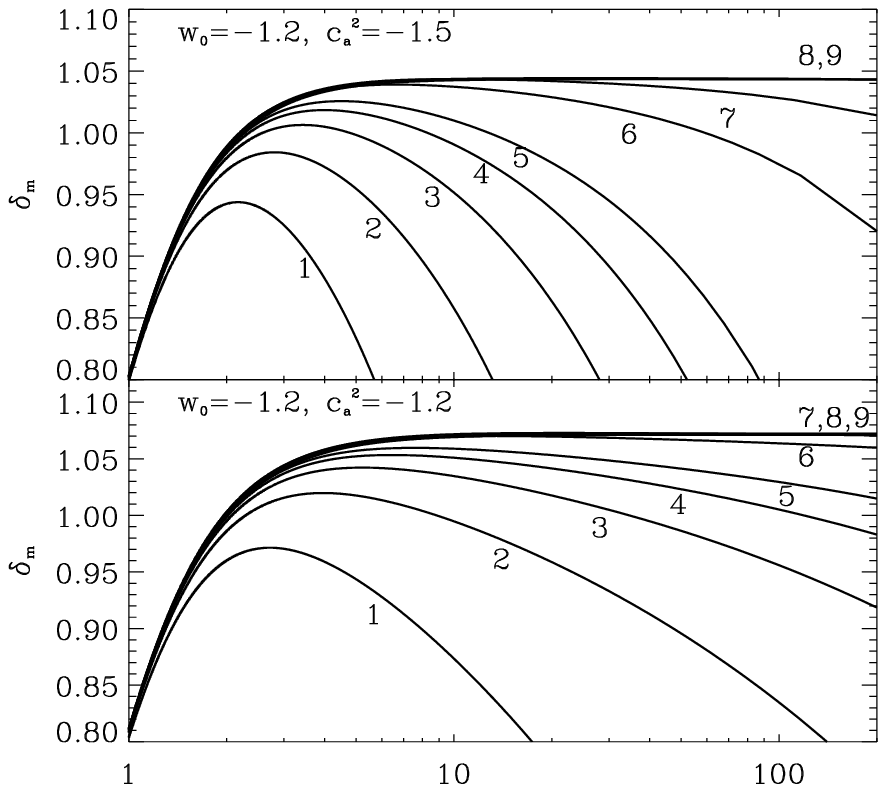}
\caption{The evolution of different Fourier amplitudes of PSF (left column) and matter (right column) density perturbations from $a=0.1$ 
to $a=200$ for models with $w_0=-1.2$, $c_a^2=-1.5$ (top panels) and $w_0=-1.2$, $c_a^2=-1.2$ (bottom panels). The rest of parameters are 
the same as in Figs.\ref{ddeb_p12_05} and \ref{fig_dm}. The different lines correspond to different wave numbers $k$ (in Mpc$^{-1}$) 
as follows: 1 - 0.0005, 2 - 0.001, 3 - 0.0015, 4 - 0.002, 5 - 0.0025, 6 - 0.005, 7 - 0.01, 8 - 0.05, 9 - 0.1 Mpc$^{-1}$. The amplitudes of 
all $k$-modes of $\delta_{de}$ are normalized to 
$\delta_{de}(k,a=0.1)=\delta_m(k,a=0.1)=0.1$ at $a=0.1$, for all $k$-modes.} 
\label{dde_p12_k}
\end{figure*}

Nevertheless the parameters of a barotropic scalar field can be constrained by  large scale structure data, since the growth rate  of matter density 
perturbations is sufficiently sensitive to them. This is illustrated in Fig. \ref{fig_dm}, where the  evolution matter density fluctuations, 
$\delta_m(a)$, is shown for models with PSF dark energy. In order to eliminate the $k$-dependence caused by the baryonic component at small scales 
and emphasize the influence of dark energy one we have normalized the amplitude of matter density perturbations to 0.1 at $a=0.1$ (free normalization). 
At this time $\rho_m/\rho_{de}\sim 1000$, $q\approx0.5$ and the amplitudes of all Fourier modes evolve essentially equally. For comparison the same 
variables for QSF, $\Lambda$CDM and the standard CDM (sCDM) models are also presented. The cosmological model with PSF can be distinguished by the 
amplitude of large scale structure inhomogeneities from  QSF at the ~10\% level and from $\Lambda$CDM at the level of a few percent for $0\le z\le 1$.

Let us analyze the evolution of linear density perturbations in the future. The first issue for clarification is the gravitational instability of PSF 
in the strongly dark energy dominated epoch. We have integrated the system of differential equations (\ref{d_de})-(\ref{h}) with initial conditions 
(\ref{init}) up to $a=200$, when expansion is already super-fast (see Fig. \ref{ah_ph}) and $\rho_{de}/\rho_m\sim 10^{8}-10^{10}$. 
The results for the PSF density perturbations are shown in Fig. \ref{dde_p12_k} (left column) in the log-norm scale for different $k$-modes 
(0.0005, 0.001, 0.0015, 0.002, 0.0025, 0.005, 0.01, 0.05, 0.1 Mpc$^{-1}$) and two expansion rates, which correspond to the models with 
$w_0=-1.2$, $c_a^2=-1.5$ (top panel) and $c_a^2=-1.2$ (bottom panel).  Their amplitudes increase slowly and the rate depends on the background 
expansion rate as well as on the wave number. In order to visualize the $k$-dependence of the growth we remove the dependence of the initial 
conditions $k$ and renormalize the amplitudes at $a=0.1$ to $\delta_{de}(k,a=0.1)=0.1=\delta_m(k,a=0.1)$, so that all $k$-modes of the PSF density 
perturbations in Fig. \ref{dde_p12_k} have the same amplitudes at $a=0.1$. The growth rate is larger for small $k$ in the range $a=0.1-10$ and it is 
practically the same for all modes at $a>10$: $\delta_{de}\propto a^{-3(1+c_a^2)/2}$. For the PSF with $w_0=-1.2$ and $c_a^2=-1.5$ the amplitude of 
the $k=0.1$ Mpc$^{-1}$ mode  increases from $a=0.1$ to $a=10$ by a factor 693, while the amplitude of the $k=0.0005$ Mpc$^{-1}$ mode increases by 10307. 
For PSF with $w_0=-1.2$ and $c_a^2=-1.2$ these numbers are 9 and 125 respectively. Since the evolution of the amplitude of the gravitational potential 
$h$ is driven by the term $\rho_m\delta_m+\rho_{de}(1+3w_{de})\delta_{de}$ (r.h.s. of eq. \ref{h}), shortly after $a=1$ the perturbations of PSF become 
important first on the largest scales and later also on smaller scales. They affect the evolution of matter density perturbations, as is shown in the 
right hand panels of Fig. \ref{dde_p12_k}. At scales with $k\ge 0.05$ Mpc$^{-1}$ (lines 8, 9 are 
superimposed in both panels) the amplitudes of matter density perturbations in the models with PSF increase from $a=1$ to $a=10$ only by a factor 
$\sim 1.3$ and freeze at this value. On these scales the difference between PSF and the $\Lambda$CDM and QSF models for the evolution of  matter density 
perturbations is inappreciable. In the $\Lambda$CDM and QSF models all $k$-modes evolve similarly to the line 9. However, on scales with 
$k<0.05$ Mpc$^{-1}$ the effect of PSF density perturbations on the evolution of matter density perturbations becomes important: the increase of PSF 
density perturbations causes the decay of matter density perturbations. The larger the scale of perturbation, the earlier its amplitude starts to 
decay\footnote{In order to visualize this effect in Fig.~\ref{dde_p12_k}, we have normalized all $k$-modes of $\delta_m$ to 0.1 at $a=0.1$.}.  

Note that this decay of matter density perturbations is caused solely by the influence of phantom scalar field perturbations, not by the super-fast 
expansion of the background (at $a\sim2$ the rates of expansion in the PSF+CDM models are close to those in $\Lambda$CDM and QSF+CDM, as it can be seen 
in Figs. \ref{wrode_ph} and \ref{ah_ph}). Excluding the effect of perturbations, the amplitudes of all $k$-modes freeze as it is shown by line 9. 
This is not the beginning of the Big Rip mentioned above, but its analog for  linear perturbations. 

\section{The best-fit parameters of PSF}\label{best-fit}

\begin{table}
\centering
 \caption{The best-fit values for cosmological parameters of PSF+CDM model and their $1\sigma$ limits from the extremal values of the N-dimensional 
distribution
determined by the MCMC technique from the combined datasets WMAP7+HST+BBN+BAO+SN SDSS SALT2 ($\mathbf{p}_1$) and WMAP7+HST+BBN+BAO+SN SDSS MLCS2k2
 ($\mathbf{p}_2$). The current Hubble parameter $H_0$ is in units km$\,$s$^{-1}\,
 $Mpc$^{-1}$. We denote the rescaled energy density of a component $X$ by $\omega_X \equiv \Omega_Xh^2$.}
 \medskip
 \begin{tabular}{ccc}
 \hline
 \hline
 Parameters&$\mathbf{p}_1$&$\mathbf{p}_2$ \medskip\\
 \hline
$\Omega_{de}$&0.72$_{-0.04}^{+0.04}$&0.69$_{-0.04}^{+0.05}$\medskip\\
$w_0$& -1.043$_{-0.24}^{+0.043}$&-1.002$_{-0.14}^{+0.002}$\medskip\\
$c_a^2$& -1.12$_{-0.50}^{+0.12}$&-1.19$_{-0.42}^{+0.19}$\medskip\\
10$\omega_b$& 0.223$_{-0.013}^{+0.016}$&0.223$_{-0.013}^{+0.014}$\medskip\\
$\omega_{cdm}$& 0.115$_{-0.010}^{+0.011}$&0.119$_{-0.010}^{+0.009}$\medskip\\
$H_0$& 70.4$_{- 3.2}^{+ 4.0}$&67.8$_{- 2.9}^{+4.2}$\medskip\\
$n_s$& 0.96$_{-0.03}^{+0.04}$&0.96$_{-0.04}^{+0.03}$\medskip\\
$\log(10^{10}A_s)$& 3.09$_{-0.09}^{+0.09}$&3.11$_{-0.11}^{+0.08}$\medskip\\
$\tau_{rei}$&0.085$_{-0.031}^{+ 0.041}$&0.086$_{-0.038}^{+0.036}$\medskip\\
 \hline
$-\log L$&3864.86&3859.30\\
 \hline
 \hline
 \end{tabular}
 \label{tab_phsf}
 \end{table}

Let us estimate the best-fit values of parameters of PSF similarly to our previous papers \cite{Novosyadlyj2010,Novosyadlyj2011,Sergijenko2011} 
devoted to QSF. We use the following datasets:
\begin{enumerate}
 \item \textit{CMB temperature fluctuations and polarization angular power spectra} from the 7-year WMAP observations (hereafter WMAP7) 
\cite{wmap7a,wmap7b};
 \item \textit{Baryon acoustic oscillations} in the space distribution of galaxies from SDSS DR7 (hereafter BAO) \cite{Percival2009};
 \item \textit{Hubble constant measurements} from HST (hereafter HST) \cite{Riess2009};
 \item \textit{Big Bang Nucleosynthesis prior} on baryon abundance (hereafter BBN) \cite{bbn,Wright2007};
 \item \textit{supernovae Ia luminosity distances} from SDSS compilation (hereafter SN SDSS) \cite{Kessler2009}, determined using the SALT2 
\cite{Guy2007} (hereafter SN SDSS SALT2) and MLCS2k2 \cite{Jha2007} methods of light curve fitting (hereafter SN SDSS MLCS2k2).
\end{enumerate}
 
In order to find the best-fit value of parameters of cosmological model with PSF and their confidence limits we perform the Markov Chain Monte-Carlo 
(MCMC) analysis for two combined datasets: WMAP7+HST+BBN+BAO+SN SDSS SALT2 and WMAP7+HST+BBN+BAO+SN SDSS MLCS2k2. We use the publicly available package 
CosmoMC \cite{cosmomc,cosmomc_source} including code CAMB \cite{camb} for the calculation of the model predictions.
\begin{table}
\centering
\caption{The best-fit values and 1$\sigma$ confidential ranges from the extremal values of N-dimensional distribution for dark energy parameters in 
$\Lambda$CDM and QSF+CDM determined by the Markov Chain Monte Carlo technique using 2 observational datasets: WMAP7+HST+BBN+BAO+SN SDSS SALT2 
($\mathbf{l}_1$, $\mathbf{q}_1$) and WMAP7+HST+BBN+BAO+SN SDSS MLCS2k2
 ($\mathbf{l}_2$, $\mathbf{q}_2$). The current Hubble parameter $H_0$ is in units km$\,$s$^{-1}\,
 $Mpc$^{-1}$.}
 \medskip
 \begin{tabular}{ccccc}
 \hline
 \hline
&$\Lambda$CDM&$\Lambda$CDM&QSF+CDM&QSF+CDM\medskip\\
 Parameters&$\mathbf{l}_1$&$\mathbf{l}_2$ &$\mathbf{q}_1$&$\mathbf{q}_2$\medskip\\
 \hline
$\Omega_{de}$&0.73$_{-0.04}^{+0.03}$&0.70$_{-0.04}^{+0.04}$&0.73$_{-0.05}^{+0.03}$&0.70$_{-0.05}^{+0.04}$ \medskip\\
$w_0$&-1 &-1& -0.996$_{-0.004}^{+0.16}$& -0.83$_{-0.17}^{+0.22}$ \medskip\\
$c_a^2$&-1 &-1& -0.022$_{-0.978}^{+0.022}$& -0.88$_{-0.12}^{+0.88}$ \medskip\\
$H_0$&70.4$_{-3.4}^{+2.9}$& 68.2$_{-3.2}^{+3.3}$&70.2$_{-4.3}^{+3.5}$&66.3$_{-3.7}^{+4.3}$\medskip\\
 \hline
$-\log L$&3864.96&3859.15&3865.01&3857.21\\
 \hline
 \hline
 \end{tabular}
 \label{tab_qsf}
 \end{table}
The results of the estimation of the PSF parameters jointly with the minimal set of cosmological parameters for the two sets of observational data 
(WMAP7+HST+BBN+BAO+SN SDSS SALT2 and WMAP7+HST+BBN+BAO+SN SDSS MLCS2k2) are presented in Table \ref{tab_phsf}. We mark the sets of best-fit parameters 
by $\mathbf{p_1}$ and $\mathbf{p_2}$. Here $\mathbf{\mathbf{p}_i}=(\Omega_{de},\,w_0,\,c_a^2,\,\Omega_b,\,\Omega_{cdm},\,H_0,\,n_s,\,A_s,\,\tau_{rei}$). 
Both SN SDSS distance moduli datasets prefer values of $w_0$ slightly lower than -1. In the past, when $a\rightarrow 0$, $w_{de}\rightarrow -1$. Hence, 
the PSFs with parameters $\mathbf{p}_1$ and $\mathbf{p}_2$ mimic a $\Lambda$-term from the Big Bang up to the current epoch, but, due to the instability 
of the value $w_{de}=-1$, even such a small difference changes drastically the future fate of the Universe: in $\Lambda$CDM model the Universe as well 
as existing structures (in principle) are time-unlimited, while in the PSF+CDM model it reaches the Big Rip 
singularity in finite time, preceded by the destruction of the structure from clusters of galaxies to elementary particles. More precisely, in the 
PSF+CDM with parameters $\mathbf{p}_1$ this happens in $\approx 152$ Gyrs, with $\mathbf{p}_2$ in $\approx 594$ Gyrs. Long before $t_{BR}$ the particle 
horizon\footnote{At the current epoch $r_p^0=14260$ Mpc in the model with $\mathbf{p}_1$ and 14170 Mpc in the model with $\mathbf{p}_2$} becomes 
$r_p^{max}\approx18710$ Mpc in model $\mathbf{p}_1$ and $\approx19200$ in model $\mathbf{p}_2$, just $\approx$1.3 times larger than the current particle 
horizon.

Let us now compare the best-fit values of cosmological parameters $\mathbf{p}_1$ and $\mathbf{p}_2$ established here with similar determinations for 
$\Lambda$CDM and QSF+CDM models. Note firstly that the subset of parameters ($w_b,\,w_{cdm},n_s,\,\,A_s,\tau_{rei}$) practically does not depend on 
the model of dark energy and SN Ia fitters, since they are determined mainly by WMAP7 data. Therefore we will compare the parameters, which depend 
on them, i. e.,  dark energy ones ($\Omega_{de}$, $w_0$, $c_a^2$) and Hubble parameter ($H_0$), and maximum of likelihoods for the same data sets. 
For the same cosmological model but with $\Lambda$ and QSF instead PSF the best-fit values of ($\Omega_{de},\, w_0,\,c_a^2,\,H_0$) are presented 
in the Table \ref{tab_qsf}. SALT2 fitting of SNe Ia light curves prefers models with lower $w_0$ than MLCS2k2 and the best-fit model with PSF has a 
slightly lower $\chi^2$ ($-\log{L}$) than $\Lambda$CDM and QSF+CDM. Although the difference is not significant, we can say that the dataset 
WMAP7+HST+BBN+BAO+SN SDSS SALT2 slightly prefers phantom models of dark energy. The dataset WMAP7+HST+BBN+BAO+SN SDSS MLCS2k2, in contrary, 
slightly prefers the quintessence model. SNe Ia distance moduli determined with use of SALT2 fitter predict also higher best-fit value of Hubble 
parameter $H_0$ than the distance moduli for the same SNe determined with MLCS2k2 fitter do. In the paper \cite{Kessler2009} the differences between 
2 methods of light curve fitting, SALT2 and MLCS2k2, are thoroughly analyzed but convincing arguments for one or the other are not given. Using them 
for the same SNe Ia samples, in~\cite{Kessler2009} it was found that the distance moduli determined with SALT2 fitter prefer lower best-fit values of 
$w_{de}=const$ than those determined with MLCS2k2. Our results support this conclusion. We have also shown in Ref.~\cite{Sergijenko2011} that the data 
on SNe Ia from the SDSS compilation with MLCS2k2 fitter allow to constrain $c_a^2$ in the quintessence range while the same data with SALT2 
fitting do not. 

Let us finally make use of the newer data on SNe Ia distance moduli from
\begin{itemize}
 \item SNLS3 compilation (hereafter SNLS3) \cite{snls3} and
 \item Union2.1 compilation (hereafter Union2.1) \cite{union}
\end{itemize}
together with data on BAO from the WiggleZ Dark Energy Survey (hereafter WiggleZ) \cite{wigglez}. The results for the combined datasets 
WMAP7+HST+BBN+BAO+WiggleZ+SNLS3 and WMAP7+HST+BBN+BAO+WiggleZ+Union2.1 are presented in Tables \ref{tab_snls} and \ref{tab_union} correspondingly.

\begin{table}
\centering
\caption{The best-fit values and 1$\sigma$ confidence ranges of the N-dimensional distribution for the dark energy parameters in QSF+CDM, 
$\Lambda$CDM and PSF+CDM determined by the Markov Chain Monte Carlo technique using the dataset WMAP7+HST+BBN+BAO+WiggleZ+SNLS3. The current 
Hubble parameter $H_0$ is in units km$\,$s$^{-1}\,
 $Mpc$^{-1}$.}
 \medskip
 \begin{tabular}{cccc}
 \hline
 \hline
 Parameters&QSF+CDM&$\Lambda$CDM&PSF+CDM\medskip\\
 \hline
$\Omega_{de}$&0.72$_{-0.04}^{+0.04}$&0.73$_{-0.04}^{+0.04}$&0.73$_{-0.04}^{+0.04}$ \medskip\\
$w_0$&-0.994$_{-0.006}^{+0.14}$&-1& -1.10$_{-0.27}^{+0.10}$\medskip\\
$c_a^2$&-0.72$_{-0.28}^{+0.72}$&-1& -1.29$_{-0.33}^{+0.29}$ \medskip\\
$H_0$&70.1$_{-4.6}^{+3.6}$&70.3$_{-3.4}^{+3.5}$&71.5$_{-4.1}^{+5.1}$\medskip\\
 \hline
$-\log L$&3947.00&3946.75&3945.98\\
 \hline
 \hline
 \end{tabular}
 \label{tab_snls}
 \end{table}
\begin{table}
\centering
\caption{The best-fit values and 1$\sigma$ confidence ranges of the N-dimensional distribution for the dark energy parameters in QSF+CDM, 
$\Lambda$CDM and PSF+CDM determined by the Markov Chain Monte Carlo technique using the observational dataset WMAP7+HST+BBN+BAO+WiggleZ+Union2.1. 
The current Hubble parameter $H_0$ is in units km$\,$s$^{-1}\,
 $Mpc$^{-1}$.}
 \medskip
 \begin{tabular}{cccc}
 \hline
 \hline
 Parameters&QSF+CDM&$\Lambda$CDM&PSF+CDM\medskip\\
 \hline
$\Omega_{de}$&0.72$_{-0.04}^{+0.03}$&0.72$_{-0.04}^{+0.04}$&0.73$_{-0.04}^{+0.03}$ \medskip\\
$w_0$&-0.995$_{-0.005}^{+0.17}$&-1& -1.13$_{-0.23}^{+0.13}$\medskip\\
$c_a^2$&-0.55$_{-0.45}^{+0.55}$&-1& -1.54$_{-0.09}^{+0.54}$ \medskip\\
$H_0$&69.7$_{-4.5}^{+3.1}$&69.8$_{-3.2}^{+3.2}$&71.4$_{-4.4}^{+4.7}$\medskip\\
 \hline
$-\log L$&3800.89&3800.76&3800.48\\
 \hline
 \hline
 \end{tabular}
 \label{tab_union}
 \end{table}
Both these combined datasets prefer phantom fields, with the best-fit values of $w_0$ and $c_a^2$ lower than in the case of SN SDSS. 
The best-fit values of $H_0$ for the phantom case are also in these cases higher. Nevertheless for these datasets the differences in 
the $-\log L$ between PSF, $\Lambda$ and QSF are still statistically insignificant.

Note also that the determination of $c_a^2$ is not sufficiently reliable for all models and datasets except for QSF and 
WMAP7+HST+BBN+BAO+SN SDSS MLCS2k2 dataset. The constraints on $w_0$ are reliable in all cases. For more details see Appendix \ref{app}.

The results of determination of cosmological parameters, especially $H_0$, $\Omega_{de}$, $w_{de}$ and $c_a^2$, presented in the 
Tables \ref{tab_phsf}-\ref{tab_qsf}, also indicate certain inconsistency or tension between fitters SALT2 and MLCS2k2 applied to the 
same SNe Ia. It was clearly highlighted and analyzed in the papers \cite{Kessler2009,Bengochea2011}, but up to now we have not decisive 
arguments for favor of one from them.

Therefore, at the current level of accuracy of cosmological data we cannot clearly distinguish between the 
nature of dark energy studied in this work -- quintessence scalar field, phantom scalar field or simply $\Lambda$.
Hopefully the data from observations which are currently under way or planned will enable us to establish the type of dark energy.

\section{Conclusion}\label{conclusions}

We have analyzed the properties of a phantom scalar field with barotropic EoS as possible dark energy candidate.
We have studied its effect on the past and future dynamics of expansion of the Universe as well as on the formation of large scale structure. 
We show that it mimics a cosmological constant (or vacuum) model of dark energy $w_{de}=-1$
 at the Big Bang, with  an EoS parameter decreasing to $w_0<-1$ at the current epoch and asymptotically approaching $c_a^2<w_0$ in the future. 
The parameters $w_0$ and $c_a^2$ together with density one $\Omega_{de}$ are free parameters of the phantom barotropic scalar field model of dark 
energy which completely define its physical variables and properties. The energy density of such a scalar field is always positive and increases 
with the expansion of the Universe. The PSF rolls up the potential with expansion, the kinetic term increases monotonically but remains always 
smaller than the potential. Only in the special case $c_a^2=w_0$, $w_{de}=const<-1$ the field starts with vanishing energy density.  The increasing 
scalar field energy density goes in hand with a super-fast expansion of the Universe, which implies a freeze in of the particle horizon and the decay 
of the event horizon. Within finite time the Universe reaches a Big Rip singularity.  

The best-fit PSF parameters have been determined jointly with all relevant cosmological parameters by the MCMC method using the data sets 
WMAP7+HST+BBN+BAO+SN SDSS SALT2 and WMAP7+HST+BBN+BAO+SN SDSS MLCS2k2 (Table \ref{tab_phsf}). From the early epoch up to now the best-fit 
PSF deviates only slightly from $\Lambda$CDM, but the future of the Universe differs significantly. In the PSF model with $\mathbf{p}_1$ 
parameters the Big Rip singularity occurs 152 Gyrs after Big Bang, while in the models with $\mathbf{p}_2$ parameters it occurs in 594 Gyrs.

The same computations have been carried out for $\Lambda$CDM and a quintessence scalar field model of dark energy and it was shown that the 
dataset WMAP7+HST+BBN+BAO+SN SDSS SALT2 slightly prefers the phantom model of dark energy, while WMAP7+HST+BBN+BAO+SN SDSS MLCS2k2 slightly 
prefers the quintessence model. However, the differences in the maximum likelihoods are statistically insignificant. The same conclusions 
apply to the datasets WMAP7+HST+BBN+BAO+WiggleZ+SNLS3 and WMAP7+HST+BBN+BAO+WiggleZ+Union2.1. We hope that more accurate future observations 
will enable us to distinguish between these models.  

\begin{acknowledgments}
This work was supported by the project of Ministry of Education and Science of Ukraine (state registration number 0110U001385), research program 
``Cosmomicrophysics'' of the National Academy of
Sciences of Ukraine (state registration number 0109U003207) and the SCOPES project No. IZ73Z0128040 of Swiss National Science Foundation. 
Authors also acknowledge the usage of CAMB and CosmoMC packages.
\end{acknowledgments} 

\appendix
\section{Estimation of the dark energy parameters: details}\label{app}

We have estimated the dark energy parameters $\Omega_{de}$, $w_0$ and $c_a^2$ jointly with the standard cosmological parameters performing a series of CosmoMC runs for different scalar fields and datasets. Each run had 8 chains converged to $R-1<0.01$.

The obtained constraints on the dark energy density parameter are reliable.

\begin{figure*}
\centerline{\includegraphics[width=\textwidth]{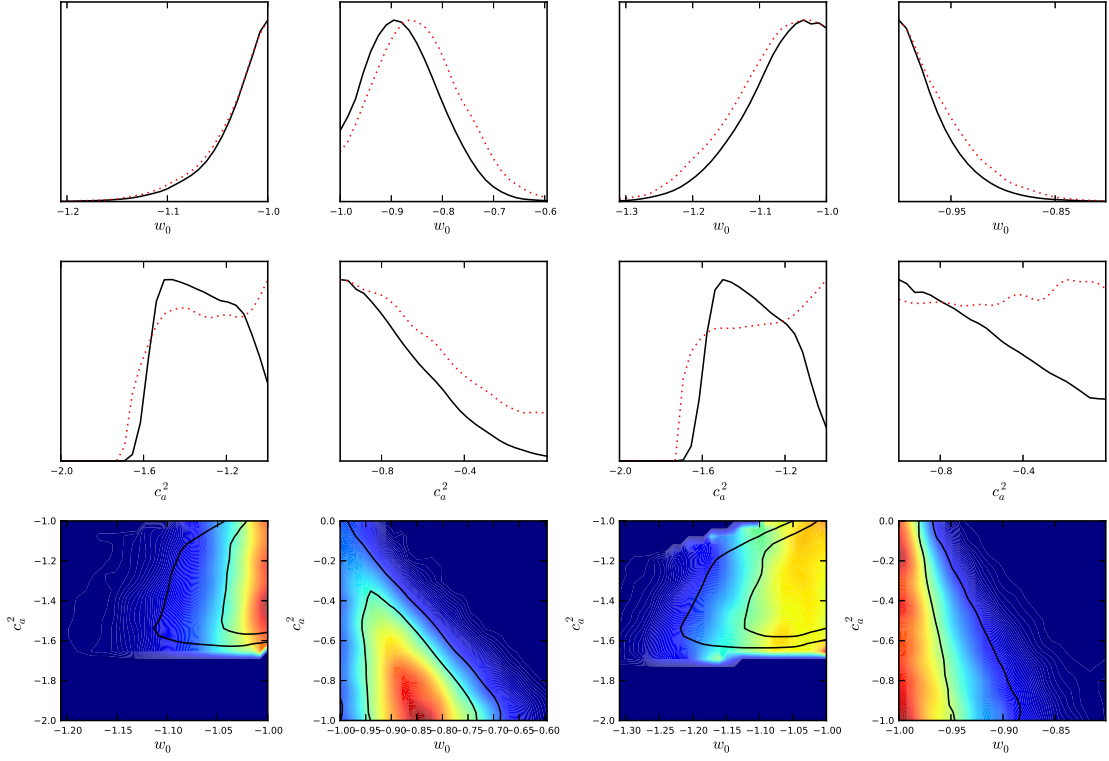}} 
\caption{One-dimensional marginalized posteriors (solid lines) and mean likelihoods (dotted lines) for $w_0$ (top panels) and $c_a^2$ (middle panels). From left to right: PSF and WMAP7+HST+BBN+BAO+SN SDSS MLCS2k2, QSF and WMAP7+HST+BBN+BAO+SN SDSS MLCS2k2, PSF and WMAP7+HST+BBN+BAO+SN SDSS SALT2, QSF and WMAP7+HST+BBN+BAO+SN SDSS SALT2. Bottom: the corresponding two-dimensional mean likelihood distributions in the plane $c_a^2-w_0$. Solid lines show the $1\sigma$ and $2\sigma$ confidence contours.}
\label{postlike_sdss}
\end{figure*}

\begin{figure*}
\centerline{\includegraphics[width=\textwidth]{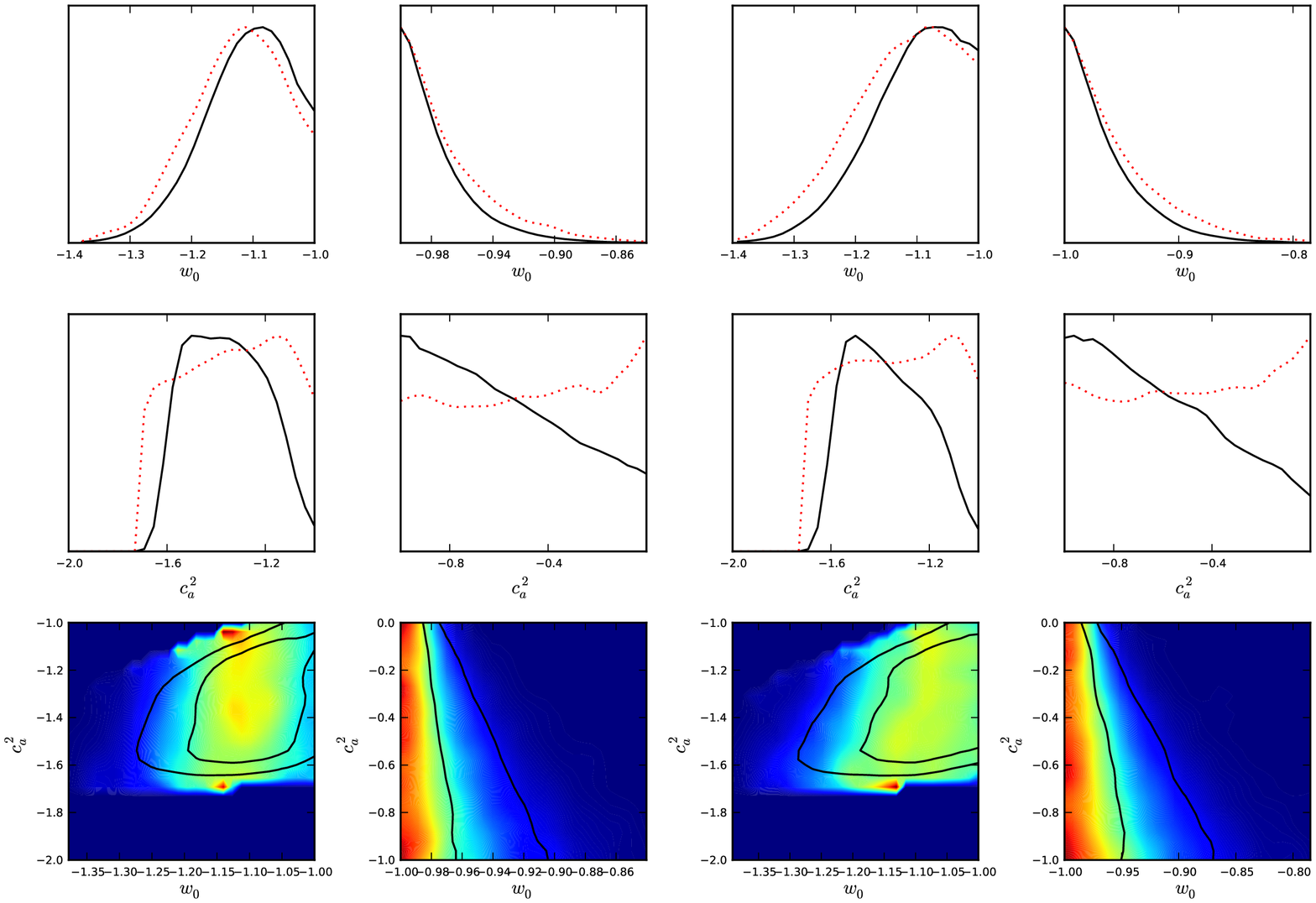}} 
\caption{One-dimensional marginalized posteriors (solid lines) and mean likelihoods (dotted lines) for $w_0$ (top panels) and $c_a^2$ (middle panels). From left to right: PSF and WMAP7+HST+BBN+BAO+WiggleZ+SNLS3, QSF and WMAP7+HST+BBN+BAO+WiggleZ+SNLS3, PSF and WMAP7+HST+BBN+BAO+WiggleZ+Union2.1, QSF and WMAP7+HST+BBN+BAO+WiggleZ+Union2.1. Bottom: the corresponding two-dimensional mean likelihood distributions in the plane $c_a^2-w_0$. Solid lines show the $1\sigma$ and $2\sigma$ confidence contours.}
\label{postlike_new}
\end{figure*}

For $w_0$ and $c_a^2$ the one-dimensional marginalized posteriors and mean likelihoods are presented in the top and middle panels of Fig. \ref{postlike_sdss} for the combined datasets WMAP7+BBN+HST+BAO+SN SDSS (MLCS2k2 and SALT2) and Fig. \ref{postlike_new} for the datasets WMAP7+BBN+HST+BAO+WiggleZ+SN SNLS3 and WMAP7+BBN+HST+BAO+WiggleZ+SN Union2.1. We see that the mean likelihoods and posteriors for $w_0$ are close to each other for both phantom and quintessence fields for all considered datasets. The shapes of curves for phantom and quintessence look like the parts of a single Gaussian cut by prior at the phantom divide $w_0=-1$ (note that the dependences in each panel are normalized to 1 at the maximum, so the absolute values of posteriors and likelihoods in the neighborhood of -1 can be close for quintessence and phantom while the normalized values are sufficiently different). The peak of Gaussian for the dataset WMAP7+BBN+HST+BAO+SN SDSS MLCS2k2 corresponds to the quintessence regime of values while 
for all other datasets to the phantom regime.

For $c_a^2$ the shapes of marginalized posteriors and mean likelihoods are significantly different for all cases except for QSF with the dataset WMAP7+HST+BBN+BAO+SN SDSS MLCS2k2 (for this case they have the shape of half-Gaussian with the center at $c_a^2=-1$). Moreover, in all cases with the mentioned exception the shapes of mean likelihoods are either asymmetric or almost flat, that is, far from a Gaussian or half-Gaussian shape. Therefore, the constraints on value of $c_a^2$ are not reliable for all cases except for QSF combined with the WMAP7+HST+BBN+BAO+SN SDSS MLCS2k2 data.

In the bottom panels of Figs. \ref{postlike_sdss} and \ref{postlike_new} the two-dimensional mean likelihood distributions in the plane $c_a^2-w_0$ are shown. The solid lines present the $1\sigma$ and $2\sigma$ confidence contours. From these plots it follows that the contours are not closed but cut by priors on the dark energy type (as well as the mean likelihood distribution). To increase the reliability of the constraints  a joint consideration of both quintessence and phantom fields is needed. This will be the topic of a separate paper.

\begin{thebibliography}{99}
\bibitem{Caldwell2002} Caldwell R.R., Phys.Lett. B {\textbf 545}, 23 (2002); arXiv:astro-ph/9908168.
\bibitem{Starobinsky2000} Starobinsky A.A., Grav. Cosmol. 6 (2000); arXiv:astro-ph/991205.
\bibitem{Perlmutter1998} Perlmutter S., Aldering G., della Valle M. et al., Nature, \textbf{391}, 51 (1998);
Perlmutter S., Aldering G., Goldhaber G. et al., Astrophys. J. \textbf{517}, 565 (1999).
\bibitem{Riess1998} Riess A.G., Filippenko A.V., Challis P. et al., Astron. J. \textbf{16}, 1009 (1998);
Schmidt B.P., Suntzeff N.B., Phillips M.M. et al., Astrophys. J. \textbf{507}, 46 (1998).
\bibitem{Alam2004} Alam U., Sahni V., Deep Saini T., Starobinsky A. A., Mon. Not. Roy. Astron. Soc. \textbf{354}, 063512 (2004).
\bibitem{Corasaniti2004} Corasaniti P.S., Kunz M., Parkinson D., Copeland E.J., Bassett B.A., Phys. Rev. D \textbf{70}, 083006 (2004);
\bibitem{Caldwell2003} Caldwell R.R., Kamionkowski M. and Weinberg N.N., Physical Review Letters {\textbf 91}, 071301 (2003).
\bibitem{Carroll2003} Carroll S.M., Hoffman M., Trodden M., Phys. Rev. D \textbf{68}, 023509 (2003);
 Singh P., Sami M., Dadhich N., Phys. Rev. D \textbf{68}, 023522 (2003); 
Cline J.M., Jeon S., Moore G.D., Phys. Rev. D \textbf{70}, 043543 (2004); 
Sami M., Toporensky A.,	Mod. Phys. Lett. A \textbf{19}, 1509 {2004}; Das S., Corasaniti P.S., Khoury J., Phys. Rev. D \textbf{73}, 083509 (2006).
\bibitem{WMAP} Spergel D. N. et. al., Astrophys. J. Suppl. Ser. \textbf{170}, 377 (2007); 
Komatsu E. et al., Astrophys. J. Suppl. Ser. \textbf{180}, 330 (2009);
 Komatsu E., et.al., Astrophys. J. Suppl. Ser. \textbf{192}, 18 (2011).
\bibitem{M-brane} Sahni V., Shtanov Yu., J. Cosmol. Astropart. Phys. \textbf{11}, 14 (2003); 
Lue A., Starkman G.D., Phys. Rev. D \textbf{70}, 101501 (2004).
\bibitem{Neupane2006} Neupane I.P., Class. Quan. Grav., \textbf{23}, 7493 (2006); 
Aref'eva I.Ya., Volovich I.V., Theor. Math. Phys. \textbf{155}, 503 (2008).
\bibitem{Elizalde2004} Elizalde E., Nojiri S., Odintsov S.D., Phys. Rev. D \textbf{70}, 043539 (2004).
\bibitem{Gannouji2006} Gannouji R., Polarski D., Ranquet A., Starobinsky A., J. Cosmol. Astropart. Phys. {\textbf 09}, 016 (2006).
\bibitem{Onemli2002} Onemli V.K., Woodard R.P., Class. Quant. Grav. \textbf{19}, 4607 (2002). 
\bibitem{Onemli2004} Onemli V.K., Woodard R.P., Phys. Rev. D \textbf{70}, 107301 (2004). 
\bibitem{Feng2005} Feng B., Wang X.L. and Zhang X.M., Phys. Lett. B \textbf{607}, 35 (2005).
\bibitem{Arkani2004} Arkani-Hamed N., Cheng H.C., Luty M.A. and Mukohyama S., JHEP \textbf{05}, 074 (2004); 
Piazza F. and Tsujikawa S., J. Cosmol. Astropart. Phys. {\textbf 07}, 004 (2004).
\bibitem{Deffayet2010} Deffayet C., Pujolas O., Sawicki I., Vikman A., J. Cosmol. Astropart. Phys. {\textbf 10}, 026 (2010).
\bibitem{S-brane} Chen C.M.,  Gal’tsov D.V., Gutperle M., Phys. Rev. D \textbf{66}, 024043 (2002); Townsend P.K., Wohlfarth M.N.R., Phys. Rev. Lett. \textbf{91}, 061302 (2003); Ohta N., Phys. Lett. B \textbf{558}, 213 (2003); Phys. Rev. Lett. \textbf{91}, 061303 (2003); Prog. Theor. Phys. \textbf{110}, 269 (2003); Int. J. Mod. Phys. A \textbf{20}, 1 (2003); Roy S., Phys. Lett. B \textbf{567}, 322 (2003).
\bibitem{Novosyadlyj2010} Novosyadlyj B., Sergijenko O., Apunevych S., Pelykh V., Phys. Rev. D \textbf{82}, 103008 (2010).
\bibitem{Novosyadlyj2011} Novosyadlyj B., Sergijenko O., Apunevych S., Journal of Physical Studies {\textbf 15}, 1901 (2011). 
\bibitem{Sergijenko2011} Sergijenko O., Durrer R., Novosyadlyj B., J. Cosmol. Astropart. Phys. {\textbf 08}, 004 (2011).
\bibitem{Ma1995} Ma C.P. and Bertschinger E., Astrophys. J. \textbf{455}, 7 (1995).
\bibitem{Durrer2008} Durrer R., The Cosmic Microwave Background, Cambridge University Press, Cambridge, 401 p. (2008).
\bibitem{camb} Lewis A., Challinor A. and Lasenby A., Astrophys. J. \textbf{538}, 473 (2000); http://camb.info.
\bibitem{class} Blas D., Lesgourgues J., Tram T., J. Cosmol. Astropart. Phys. \textbf{07}, 034 (2011).
\bibitem{Garriga1999} Garriga J., Mukhanov V.F., Phys. Lett. B \textbf{458}, 219 (1999).
\bibitem{wmap7a} Jarosik N., Bennett C.L., Dunkley J., Gold B., Greason M.R. et al., Astrophys. J. Suppl. \textbf{192}, 14 (2011).
\bibitem{wmap7b} Larson D., Dunkley J., Hinshaw G., Komatsu E., Nolta N.R. et al., Astrophys. J. Suppl. \textbf{192}, 16 (2011).
\bibitem{Percival2009} Percival W.J., Reid B.A., Eisenstein D.J., Bahcall N.A., Budavari T. et al., Mon. Not. Roy. Astron. Soc. \textbf{401}, 2148 (2010).
\bibitem{Riess2009} Riess A.G., Macri L., Casertano S., Sosey M. et al., Astrophys. J. \textbf{699}, 539 (2009).
\bibitem{bbn} Steigman G., Ann. Rev. Nucl. Part. Sc. \textbf{57}, 463 (2007).
\bibitem{Wright2007} Wright E.L., Astrophys. J. \textbf{664}, 633 (2007).
\bibitem{Guy2007} Guy J., Astier P., Baumont S., Hardin D., Pain R. et al., Astron. \&  Astrophys. \textbf{466}, 11 (2007).
\bibitem{Jha2007} Jha S., Riess A.G. and Kirshner R.P.,  Astrophys. J. \textbf{659}, 122 (2007).
\bibitem{cosmomc} Lewis A. and Bridle S., Phys. Rev. D \textbf{66}, 103511 (2002).
\bibitem{cosmomc_source} http://cosmologist.info/cosmomc
\bibitem{Kessler2009} Kessler R., Becker A.C., Cinabro D., Vanderplas J., Frieman J.A. et al., Astrophys. J. Suppl. \textbf{185}, 32 (2009).
\bibitem{snls3} Sullivan M., Guy J., Conley A., Regnault N., Astier P. et al., Astrophys. J. \textbf{737}, 102 (2011).
\bibitem{union} Suzuki N., Rubin D., Lidman C., Aldering G., Amanullah R. et al., Astrophys. J. \textbf{746}, 85 (2012). 
\bibitem{wigglez} Blake C., Kazin E., Beutler F., Davis T., Parkinson D. et al., MNRAS \textbf{418}, 1707 (2011).
\bibitem{Bengochea2011} Bengochea G.R., Supernova light-curve fitters and dark energy, Phys. Lett. B {\textbf 696}, 5 (2011).

\end{thebibliography}
\end{document}